\documentclass[12pt]{iopart}
\usepackage{graphicx,color,bm}      
\usepackage{hyperref}
\usepackage{amssymb}
\definecolor{clr}{rgb}{0, 0.25, 0.75}
\hypersetup{
  colorlinks=true,
  linkcolor=clr,
  citecolor=clr,
  filecolor=magenta,
  urlcolor=clr,
  pdfpagemode=UseThumbs,
  pdfstartview=FitH,
}
\usepackage{cite}
\usepackage{braket}

\begin{document}
\topical{Measurement-Induced Entanglement Phase Transition in Free Fermion Systems}
\author{Han-Ze Li$^{1}$, Jian-Xin Zhong$^1$, Xue-Jia Yu$^{2,3}$}

\address{$^1$ Institute for Quantum Science and Technology, Shanghai University, Shanghai 200444, China}
\address{$^2$ Department of Physics, Fuzhou University,
Fuzhou 350116, Fujian, China}
\address{$^3$ Fujian Key Laboratory Engineering, Fuzhou University, Fuzhou, Fujian 350108, China}
\ead{xuajiayu@fzu.edu.cn}

\begin{abstract}

Measurement-induced entanglement phase transitions (MIET) highlight how local measurements drive quantum systems between area-law and volume-law entangled states. This review surveys MIET in free fermion models, focusing on how unitary hopping competes with measurement-induced non-unitarity. We discuss controversies regarding the existence of MIET in one dimension, the impact of non-Hermitian skin effects, and potential experimental platforms. We conclude with open challenges, including feedback control and higher-dimensional extensions.

\end{abstract}


\maketitle

\section{Introduction}

Understanding how quantum entanglement evolves under the influence of measurements has emerged as a central theme in contemporary quantum many-body physics. \textit{MIETs}~\cite{Fisher2023,MIETxt1,MIETxt2,MIETxt3,MIETxt4,MIETxt5,MIETxt6,MIETxt7,MIETxt8,MIETsd1,MIETsd2,MIETsd3,ls_post1,xzy1_monitored,Tony1,Tony2,Tony3,Adani2024}, wherein the entanglement entropy of a quantum system exhibits a sharp transition as a function of the measurement strength or rate, offers a compelling example of how quantum information and non-unitary dynamics can interplay to yield new dynamical phase transitions.

While much of the early exploration of MIETs was conducted within the framework of hybrid quantum circuits~\cite{Fisher2023} composed of unitary gates and stochastic measurements, growing attention has been paid to more physically grounded models. Among them, free fermion systems serve as a minimal yet nontrivial platform for investigating measurement-induced physics. These systems are not only analytically tractable but also experimentally relevant.

Additionally, what makes free fermion systems particularly intriguing is their \textit{Gaussianity}-preserving structure under both unitary and certain classes of monitored evolution. This feature allows for efficient classical simulation of entanglement dynamics even in the presence of measurements, thereby enabling large-scale numerical studies and analytical insights. Furthermore, recent developments in quantum trajectory theory~\cite{Rivas2012} and non-Hermitian quantum mechanics~\cite{nhbook,Ashida_2020,non-hermitian} have revealed that effective non-Hermitian dynamics can be reinterpreted as generating no-jump measurements evolution. This reinterpretation has catalyzed a wave of studies exploring jump-free, yet non-Hermitian, analogues of MIET.

This review provides a comprehensive overview of recent developments in MIET in free fermion systems, with an emphasis on both \textit{stochastic} (quantum-jump-based) and \textit{deterministic} (non-Hermitian) evolutions. We begin with a brief introduction to monitored dynamics and quantum trajectory formulations, followed by a detailed account of MIET in locally and globally monitored free fermion models. We then explore the steady-state entanglement phase transitions driven by effective non-Hermitian evolutions, with a special focus on the non-Hermitian skin effect. We conclude with a summary of experimental progress and open questions that are expected to shape future developments in this rapidly evolving field.


\section{Monitored Dynamics}\label{sec:2}
Before diving into MIET in monitored free fermion systems, let us briefly review several key elements of monitoring dynamics as a warm-up. Essentially, MIET describes the \textit{hybrid} evolution process of a quantum system that is repeatedly measured while undergoing its intrinsic unitary evolution. The introduction of measurements renders the system open, making the conventional unitary evolution theory inapplicable for describing this hybrid evolution. In this generalized framework, the quantum system can be characterized either by Hamiltonian evolution or by a trotterized quantum circuit of continuous-time gates; both approaches are valid. Furthermore, different measurement protocols can steer the system toward distinct evolutionary outcomes. Additionally, it is important to note that this review focuses on the hybrid evolution of free fermion systems; for a review of monitoring quantum circuits, please refer to Ref.~\cite{Fisher2023}.

\subsection{From Projective to Weak Measurements}
Consider projective measurements of an observable $O=\sum_{y} y\,\Pi_y$, with projectors $\{\Pi_y\}$. For later convenience we denote the corresponding measurement operators by the same index set, $\{M_y\}$, and they satisfy the completeness relation $\sum_y M_y^\dagger M_y = I$. In the projective limit one simply has $M_y=\Pi_y$, so upon obtaining outcome $y$ the state collapses to $\Pi_y\ket{\psi}$.

However, instead of an idealized projective measurement, there is an approach in which the measurement device takes a finite amount of time to perform the measurement. This method, more feasible for experimental setups, is known as a weak measurement. Its core idea is to use a weak coupling between the system and the measurement apparatus so that each individual measurement only slightly perturbs the system's state while still gathering partial information. To formalize this, we introduce a continuous measurement outcome $x$ and construct a series of Kraus operators 
\begin{equation}
    M_y(x) = \left( \frac{1}{2\pi\sigma^2} \right)^{1/4} \exp{\left[ -\frac{(x-\lambda y)^2}{4\sigma^2} \right]},
\end{equation}
where, $\sigma$ describes the width of the probe initial wave function. These operators satisfy the continuous completeness relation $\int dx M_y^\dagger(x)M_y(x)=I$. For a single measurement on an initial state $\ket{\psi}$, the probability density of obtaining the result $x$ is given by $p(x) = \bra{\psi}M_y^\dagger(x)M_y(x)\ket{\psi}$, and the normalized post-measurement state is updated to
\begin{equation}
    \ket{\psi(x)} = \frac{M_y(x)\ket{\psi}}{\sqrt{p(x)}}.
\end{equation}
The parameters $\lambda$ and $\sigma$ govern the nature of the measurement:
\begin{itemize}
    \item In the \emph{weak} measurement regime ($\lambda\ll 1$ or $\sigma\gg 1$) the exponential factor varies slowly with $y$, so each measurement only slightly perturbs the system and provides limited information about the eigenvalue $y$.
    \item In the \emph{strong} measurement limit ($\lambda=1$ and $\sigma\to 0$), the Gaussian function approaches a Dirac delta function $\lim_{\sigma \to 0} ( {1}/\sqrt{2\pi\sigma^2} ) \exp[-{(x - \gamma y)^2}/{2\sigma^2}] = \delta(x - y)$.
    Here, the outcome $x$ is sharply peaked around $x=y$, and the measurement operators become proportional to the projectors $M_y(x)\sim \delta(x-y)=\Pi_y$, thus recovering projective measurements.
\end{itemize}
By tuning $\lambda$ and $\sigma$, one can smoothly interpolate between weak and strong measurements, with $\lambda$ controlling the effective scale of the measurement outcome and $\sigma$ determining the resolution of the apparatus.



\subsection{Stochastic Schr\"odinger Equation}\label{sse}
To track the evolution of a monitored quantum system at the pure state level and capture the randomness introduced by measurements, one often employs the quantum trajectory method, that is, the Stochastic schr\"odinger equation. This approach is equivalent to the master equation but reflects the uncertainty introduced by the measurement process at the state-vector level. 

In Kraus representation, the evolution of the system over a very short time interval $dt$ is represented by a set of Kraus operators $\{M_{\alpha}\}$, so that the density matrix evolves as
\begin{equation}
    \rho(t+dt)=\sum_{\alpha} M_{\alpha}(dt)\rho(t)M^\dagger_{\alpha}(dt).
\end{equation}
These operators can be divided into two categories:
\begin{equation}
    M_0 = I - iH_{\rm eff}\,dt,  \\
    M_k = \sqrt{\gamma\,dt}\,L_k,
\end{equation}
Here, $H_{\rm eff}=H-\frac{i\gamma}{2}\sum_{k\geq 1}L^\dagger_kL_k$ is the non-Hermitian effective Hamiltonian and $L_k$ is the jump operator. $M_0$ and $M_k$ denote the no-jump operator (sometimes referred to as the non-Hermitian evolution operator) and the jump operator, respectively. In the discrete picture, during each small time step $dt$, the system either evolves without a jump or experiences a single jump. The probability for a jump is determined by a corresponding expression 
\begin{equation}
    p_{k\geq1}=\bra{\psi(t)}M^\dagger_k M_k\ket{\psi(t)}\approx \gamma dt\bra{\psi(t)}L^\dagger_k L_k\ket{\psi(t)}.
\end{equation}
If a jump occurs, the state vector is updated according to the jump operator, i.e.,
\begin{equation}
    \ket{\psi(t+dt)}=\frac{M_k\ket{\psi(t)}}{\sqrt{p_{k\geq 1}}},
\end{equation}
and if no jump occurs, the state vector evolves under the no-jump operator:
\begin{equation}
    \ket{\psi(t+dt)}=\frac{M_0\ket{\psi(t)}}{\sqrt{p_0}},
\end{equation}
where $p_0=1-\sum_{k\geq 1}p_{k}$ is the probability for no jump. This framework effectively captures the stochastic nature of the evolution in a monitored quantum system.

In the continuous limit $dt\rightarrow 0$, the evolution process can be described by an It\^o stochastic differential equation. One of the most common formulations is known as the \textit{quantum jump unraveling}. In this approach, the state evolves continuously under a non-Hermitian effective Hamiltonian until a quantum jump occurs. The probability of a jump in an infinitesimal time interval is determined by the jump operator, and when a jump occurs, the state is updated according to that operator. Between jumps, the state follows a deterministic (but non-unitary) evolution. Formally, the quantum jump stochastic Schr\"odinger equation can be written as
\begin{equation}
    d\ket{\psi(t)} = -iH_{\rm eff}\ket{\psi(t)}dt+\sum_{k\geq1}\left(\frac{L_k\ket{\psi(t)}}{||L_k\ket{\psi(t)}||}-\ket{\psi(t)}\right)dN_k(t),\label{Qj}
\end{equation}
where $dN_k(t)$ is the set of Poisson stochastic increment satisfying $\mathbb{E}[dN_k(t)]=\gamma dt\bra{\psi(t)}M^\dagger_k M_k\ket{\psi(t)}$. The evolution is smooth most of the time, driven by $H_{\rm eff}$, while jump or click events cause the state to undergo sudden, discrete transitions.

In certain cases, the jump process can be approximated by a continuous noise process, leading to the so-called \textit{quantum state diffusion}~\cite{Ashida_2020} formulation, whose expression is given by
\begin{equation}
    d|\psi(t)\rangle
    = -\,i\,H_{\mathrm{eff}}\,|\psi(t)\rangle\,dt
    \;+\;
    \sum_{k}\bigl(L_k - \langle L_k\rangle\bigr)\,|\psi(t)\rangle\,dW_k(t),\label{QSD}
\end{equation}
where 
\begin{equation}
    H_{\mathrm{eff}}
    = H
    - \frac{i\gamma}{2}\,\sum_{k}\,
    \Bigl(
      L_k^\dagger L_k
      \;-\; 2\,\langle L_k^\dagger\rangle\,L_k
      \;+\; |\langle L_k\rangle|^2
    \Bigr),
\end{equation}
and $\langle\cdot\rangle=\bra{\psi(t)}\cdot\ket{\psi(t)}$. In this formulation, $dW_k(t)$ denotes independent Wiener increments that satisfy $\mathbb{E}[dW_k(t)]=0$, $\mathbb{E}[dW_k(t)dW_j(t)]=\delta_{kj}dt$. In this approach, the deterministic component is consistent with the no-jump evolution, while the stochastic component captures the perturbations introduced by environmental noise or measurement uncertainty.

In general, the quantum jump stochastic schr\"odinger Eq.~(\ref{Qj}) is well-suited for describing discrete measurement events, while the quantum state diffusion stochastic schr\"odinger Eq.~(\ref{QSD}) is more appropriate for capturing the system's behavior under continuous weak measurements. However, regardless of whether one employs the quantum jump or quantum state diffusion formulation, averaging over a large number of trajectories yields the same Lindbladian density matrix evolution. 

Moreover, one of the most fundamental probes of entanglement dynamics is the \textit{von Neumann entropy}, defined as
\begin{equation}
    S_A = -{\rm Tr}(\rho_A \ln{\rho_A}),
\end{equation}
where $\rho_{A} = {\rm Tr}_{\bar{A}} \ket{\Psi}\bra{\Psi}$ is the reduced density matrix obtained by tracing out the complementary subsystem $\bar{A}$. The von Neumann entropy exhibits several notable properties; in particular, it provides a quantitative measure of entanglement in many-body systems. For a pure state, the entropy of a subsystem precisely quantifies the amount of entanglement between that subsystem and the remainder of the system.

\section{MIET in Monitored Free Fermion Systems}\label{sec:3}
To begin with, we briefly introduce the general properties of monitored free-fermion systems. Consider a free-fermion system whose dynamics are generated by a quadratic
Hamiltonian
\begin{equation}
  H=\sum_{i,j} c_i^{\dagger} h_{ij} c_j\;=\;\frac{1}{2}\,\mathbf{C}^{\dagger} h \mathbf{C},
  \qquad
  \mathbf{C}^{\mathsf{T}}=(c_1,\ldots,c_L,c_1^{\dagger},\ldots,c_L^{\dagger}),
\end{equation}
where $h$ is a real, antisymmetric Bogoliubov-de-Gennes matrix.
Unitary evolution under $H$ preserves Gaussianity: if the initial state
$\rho(0)$ is Gaussian, then the time-evolved state
$\rho(t)=e^{-iHt}\rho(0)e^{iHt}$ remains Gaussian and is fully determined by
its two-point correlation matrix
$G_{ij}(t) = \mathrm{Tr}[\rho(t)\, C_i^\dagger C_j]$. Now introduce a set of measurement operators $\{M_y\}$
that \emph{commute} with the Hamiltonian,
$[M_y,H]=0\;\forall\,y$.
Typical examples include
(i) strong projectors onto eigen-mode occupations
$N_\alpha=\tilde c_\alpha^{\dagger}\tilde c_\alpha$ and
(ii) weak Gaussian POVMs, $M_y \;=\;
  (2\pi\sigma^{2})^{-L/2}\,
  \exp{[-\!\sum_{\alpha}(N_\alpha-y_\alpha)^{2}/4\sigma^{2}]},$ whose width $\sigma$ tunes the measurement strength.
Because $M_y$ and $H$ share the same eigenbasis, both the post-measurement
conditional state
$\rho_y=M_y\rho M_y^{\dagger}/\mathrm{Tr}[M_y\rho M_y^{\dagger}]$
and the averaged state
$\sum_y M_y\rho M_y^{\dagger}$
remain Gaussian.
At the level of correlations this amounts to an affine map $  G \longmapsto G'=RGR^{\mathsf{T}} + Q$, with an orthogonal matrix $R$ and symmetric matrix $Q$ fixed solely by $\sigma$.
Taking the continuous-monitoring limit
one obtains a closed equation $  {\mathrm{d}G}/{\mathrm{d}t} = -i\,[h,G] -\sigma^{-2}\delta t(G - DGD),$
where $D$ projects onto mode-occupation space.
Hence any measurement that commutes with $H$ preserves the Gaussian
structure, ensures Wick’s theorem remains valid, and allows all entanglement
and coherence measures to be extracted from $G$ alone, providing a
self-contained and efficient framework for analysing MIET in free-fermion systems.

Before we proceed, it is useful to shortly review the fundamental concepts of MIET\footnote{We want to emphasize once again that this review focuses on free fermion systems; for developments on MIET in quantum random circuits, please refer to Ref.~\cite{Fisher2023}.}. Initially, the discussion of MIET was introduced in the context of random quantum circuits with local measurements and no conservation laws. In this setting, when the measurement strength (or rate) exceeds a critical threshold, the system’s steady-state entanglement follows a low-entanglement area law, whereas below the critical rate it transitions to a volume-law steady state. At the critical point, a logarithmic form of steady-state entanglement emerges, whose universality class can be captured by a percolation phase transition~\cite{Fisher2023,skinner}. 

However, from the perspective of physical implementation, the most natural and simplest model for discussing MIET is monitoring free fermion systems. In these systems, the entanglement bath is provided solely by coherent hopping, while local measurements on the free fermions ensure that the monitored dynamical evolution preserves the Gaussian state structure. This guarantees that such a model remains efficiently simulable even for large system sizes. Ref.~\cite{Bernard_2018} was the pioneer of consider continuous measurements on the free fermions. Than, Ref.~\cite{cao} was explored in a free fermion Hamiltonian system that preserves $U(1)$-symmetry of the system. In this study, a quasi-particle pair approach for entanglement in integrable systems was proposed based on generalized hydrodynamics. By employing an ansatz for the collapse of quasi-particle pairs and combining it with large-scale numerical simulations, the work quantitatively demonstrated that the volume-law steady-state entanglement structure arising from unitary dynamics can be disrupted by continuous measurements at any strength (rate). However, the understanding of MIET in locally monitored free fermion systems is still far from complete. A natural question arises: does MIET exist in these systems? If it does, what are its critical properties, and which universality class does it belong to? In fact, there are two competing viewpoints regarding the existence of MIET in the thermodynamic limit for such monitoring free fermion systems.

\begin{figure}[tb]
\begin{center}
\includegraphics[width=13.5cm]{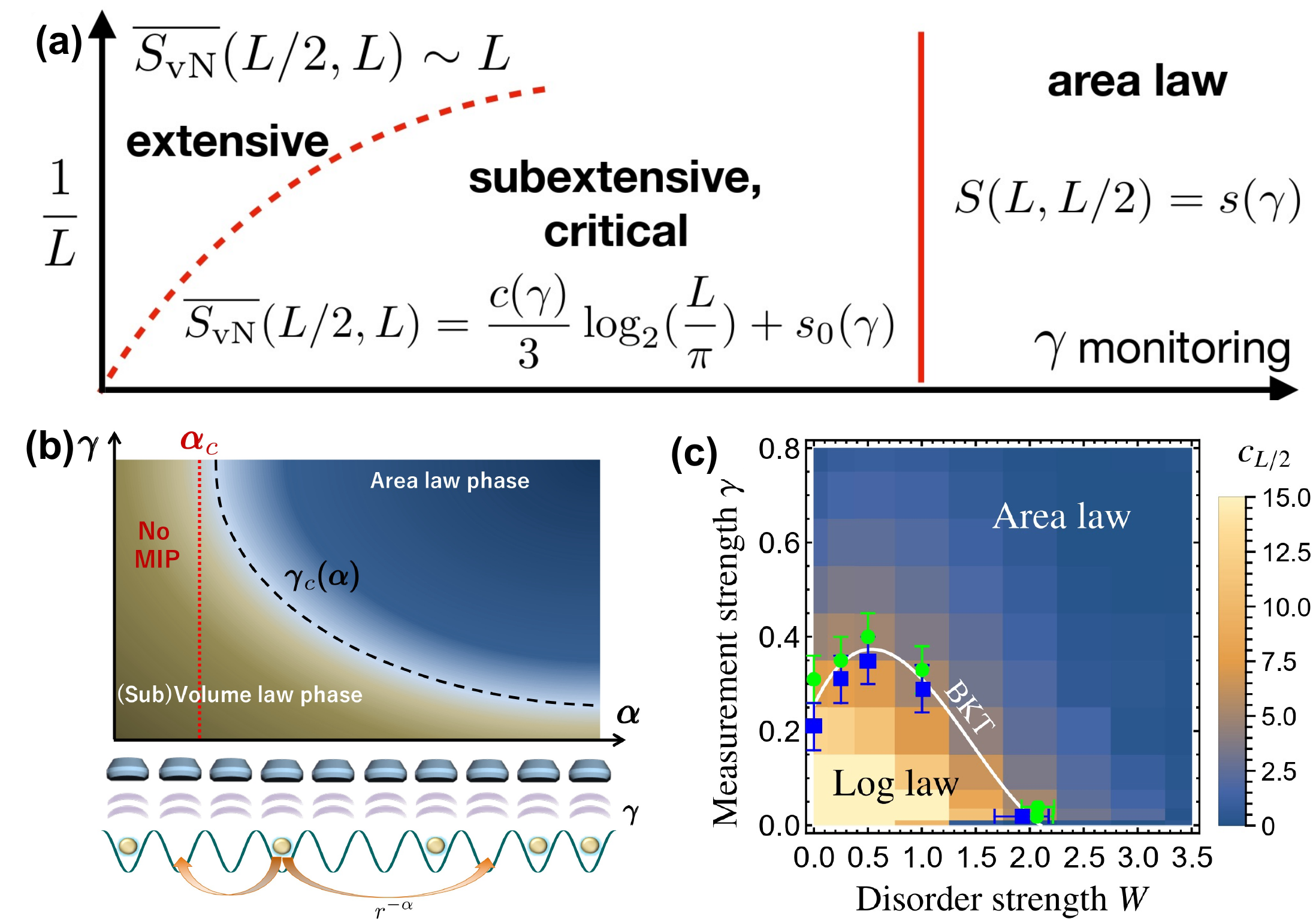}
\end{center}
\caption{\textit{Entanglement steady state phase diagram in different systems.} Panel (a) corresponds to Ref.~\cite{MIET2021}, where the phase diagram is studied in a free-fermion hopping chain of length $L$ subject to continuous monitoring at a dimensionless rate $\gamma$. Panel (b) presents the phase diagram of monitored free-fermion chains with long-range couplings, reprinted from Ref.~\cite{Minato_2022}. Panel (c) shows the phase diagram obtained in Ref.~\cite{disorderedfreefermion} for disordered, monitored free-fermion chains.}
\label{fig:free}
\end{figure}

On the one hand, evidence indicates that monitored free fermion 1D systems exhibit a measurement-induced Berezinskii-Kosterlitz-Thouless (BKT) type phase transition. In this context, Ref.~\cite{effectivetheory} developed an $n$-replica Keldysh field theory approach, and its statistical analysis revealed a nontrivial BKT transition driven by measurement degrees of freedom in a system of spinless Dirac fermions. The study further clarified that while entanglement entropy is a hallmark of MIET, correlation functions can also provide similar information. Additionally, Ref.~\cite{MIET2021} provided numerical evidence showing that continuously measured free fermion systems can undergo a BKT-type phase transition. Their numerical analysis linked the gapless phase with conformal invariance and the logarithmic scaling of entanglement entropy, see panel (a) of Fig.~\ref{fig:free}. A series of intriguing variations followed. Refs.~\cite{Minato_2022,darklong} examined the fate of the steady-state entanglement MIET in monitoring free fermion chains with long-range coupling. They also studied a fermionic Hamiltonian model with long-range hopping and found that systems with long-range coupling exhibit behavior similar to short-range MIET systems, but only when the long-range decay exponent $\alpha >  1.5$; otherwise, MIET is absent, see panel (b) of Fig.~\ref{fig:free}. Furthermore, Ref.~\cite{disorderedfreefermion} investigated whether introducing a disorder localization potential in a monitored free fermion chain would destroy the BKT universality. Their findings indicate that the competition between measurement and disorder leads to rich steady-state entanglement structures, with the BKT transition remaining robust against disorder, see panel (c) of Fig.~\ref{fig:free}. Refs.~\cite{boudary,boundary2} focused on the entanglement dynamics in boundary-driven monitored free fermion systems, exploring the impact of continuous measurements on open quantum systems. Their work revealed that measurement plays a dual role in affecting late-time entanglement, inducing critical behavior, and altering quantum transport properties. In addition, studies on driven many-body systems have demonstrated that a high measurement rate can induce local correlations and anomalous transport. Regarding the fate of periodically driven monitored free fermion systems, Ref.~\cite{MIETperiodically} performed a renormalization group analysis and numerical validation of the non-Hermitian sine-Gordon model introduced in Ref.~\cite{effectivetheory} under the high measurement rate limit. The results indicate that the BKT phase transition persists in these periodically driven systems, although its occurrence depends on the specific driving protocol. These variations indicate that the BKT universality class found in monitored free fermion 1D systems is robust against long-range hopping, disorder quenches, and periodic driving. Essentially, this robustness emerges from the competition between measurement-induced non-unitary evolution and the system's intrinsic unitary dynamics. However, it is well known that changing the dimensionality of a system can significantly affect its critical behavior, scaling laws, and renormalization group currents. Ref.~\cite{twoD} explored continuously measured 2D free fermion systems and discovered that the localization-delocalization MIET in these systems can be described by a unified binary framework involving the $SU(R)$ nonlinear sigma model and the $SU(2R)$ non-Hermitian Hubbard model. Moreover, the wavefunctions exhibit conformal invariance at specific measurement strengths, exceeding the predictions of Anderson localization theory. In other words, these findings suggest that high-dimensional MIET in free fermion systems may feature richer universality classes and novel phase transition mechanisms. Additionally, Ref.~\cite{xzy_topo} presents a classification of topological insulators and superconductors, analogous to the periodic table of topological phases in open quantum systems. It provides a systematic framework for exploring how monitored free fermion systems exhibit topological features across different symmetry classes and spacetime dimensions.

On the other hand, recent studies have suggested that MIET in monitored free fermion 1D systems may not exist in the thermodynamic limit. Specifically, Ref.~\cite{projective} investigated the competition between unitary dynamics and projective measurements in these systems. Their main finding is that the volume-law phase disappears at any measurement rate, leaving no MIET. Numerical results indicate that in the thermodynamic limit, the average steady-state entanglement entropy exhibits a single transition from volume-law to area-law, irrespective of the measurement rate. Although significant finite-size effects in free fermion systems lead to a residual logarithmic scaling, Ref.~\cite{nonlinear} further developed a replica Keldysh path-integral method. By deriving a nonlinear sigma model as an effective field theory and combining it with numerical results for time density matrix correlation functions, they demonstrated that monitored free fermion systems eventually tend toward a localized area-law in the thermodynamic limit. This supports the conclusion that an entanglement transition corresponding to MIET is absent in the thermodynamic limit. Ref.~\cite{PL} presents an exact analytical solution for the effective continuum field theory of a monitored free fermion system with conserved charge, and also demonstrates the absence of MIET in such charge-conserving free fermion systems.

Separately, modifying the quantum jump protocol offers new insights into MIET dynamics. Ref.~\cite{roleofQJ} introduced a novel framework to study the statistical distribution of entanglement entropy gain and loss before and after quantum jumps. Within this framework, it was demonstrated that under weak measurement conditions, there exists a significant difference between a monitoring free fermion chain with $U(1)$ symmetry and an Ising chain, even though the two can be connected via a Jordan-Wigner transformation. In the case of the free fermion chain, the normalized non-Hermitian Hamiltonian does not contribute significantly to the scaling of entanglement entropy, whereas such normalization is absent in the Ising chain. Moreover, recent work has drawn considerable attention to a class of general monitoring free fermion 1D systems that incorporate a feedback operation~\cite{wyp,Feng_2023,lzc,fx,huang}. In these systems, a specific feedback operator links quantum jump dynamics with non-Hermitian effective Hamiltonian dynamics on a time-scale basis rather than in the steady-state limit. The primary consequence is that the system dynamically exhibits metastable states and novel dynamical phase transitions, phenomena that are distinct from those observed in conventional MIET in monitored free fermion 1D systems. Additionally, Ref.~\cite{levy} examined the entanglement structure of a locally monitored free fermion chain and found that local disturbances can lead to an anomalously slow growth of entanglement entropy, eventually reaching a volume-law phase. This finding highlights the key role of quantum jumps in generating entanglement: although quantum jumps reduce entanglement entropy, they also rearrange nearby many-body states. This insight further clarifies the interplay between quantum jump dynamics and non-Hermitian evolution in monitoring free fermion systems.

\section{MIET in non-Hermitian Free Fermion Systems}\label{sec:4}

In the previous section, we have seen that monitoring free fermion systems may exhibit MIET, although this remains a subject of debate. Fundamentally, the MIET in these systems arises from the dynamic competition between the intrinsic unitary dynamics and the non-unitary dynamics introduced by measurements. In the language of quantum trajectories, measurements not only introduce dynamical randomness but also rewrite the evolution Hamiltonian. This naturally raises the questions: Is the dynamical randomness introduced by measurements necessary for the emergence of a steady-state entanglement phase transition? In other words, can a pure effective evolution also lead to a steady-state entanglement phase transition? 

To address this question, we will focus on non-Hermitian evolution. In measurement quantum systems, it is always possible to post-select the trajectories corresponding to no-click events and consider only the pure non-Hermitian effective evolution, although implementing this in the laboratory poses an exponentially difficult challenge. This demonstrates that the specific form of the non-Hermitian effective Hamiltonian is entirely determined by the measurement operators. Consequently, effective non-Hermitian Hamiltonian constructed with appropriate measurement operators can, in theory, contribute non-unitary dynamics that compete with the intrinsic unitary dynamics, thereby leading to steady-state entanglement phase transitions. It is worth noting that non-Hermitian quantum mechanics has a long history, dating back to studies on disordered models of non-interacting electrons, non-unitary conformal field theories, and systems with parity-time ($\mathcal{PT}$) symmetry~\cite{bender1,bender2}. In recent years, non-Hermitian physics has attracted widespread attention for its unique phenomena~\cite{Ashida_2020,wang1,wang2,wang3,wang4,wang5,ch1,ch2,ch3,ch4,ch5,ch6,ch7,ch8,ch9,ch10,ch11,ch12,ch13,ch14,ep,sz1,sz2,sz3,sz4,sz5,Li1,Li2,gzx,xjy,nhskin,lzc2,lsz,zhou1,zhou2,pointgap,yxj1,yxj2,chx,lqy} that extend beyond those observed in Hermitian systems, while also differing from noisy quantum circuits~\cite{ls1,ls2,ls3}. In this section, we will review the rich landscape of steady-state entanglement in monitoring free fermion systems under pure non-Hermitian effective evolutions,

\begin{equation}
    d\ket{\psi(t)} = -iH_{\rm eff}\ket{\psi(t)}dt.
\end{equation}

\subsection{Entanglement Transitions in Measurement-Induced Non-Hermitian Skin Effects}

\begin{figure}[tb]
\begin{center}
\includegraphics[width=13.5cm]{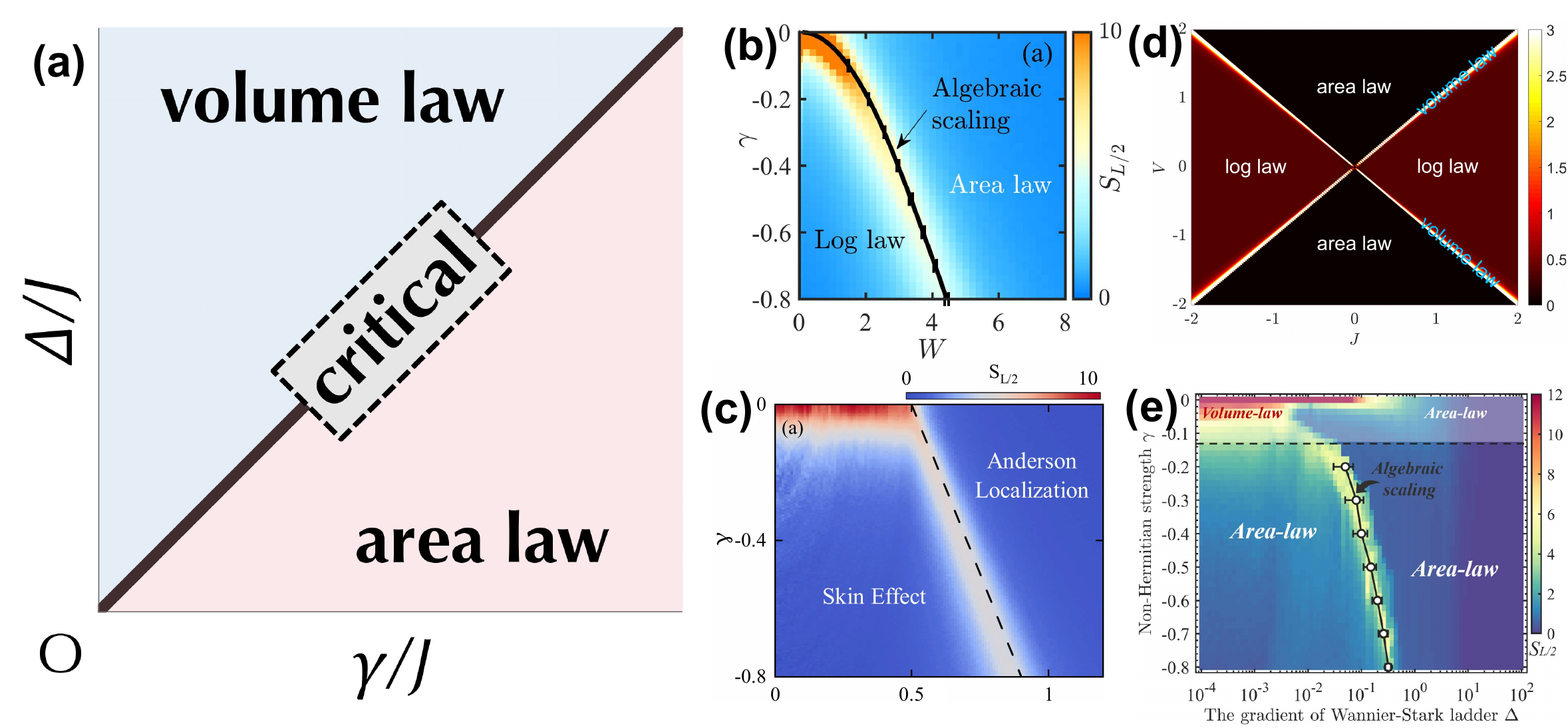}
\end{center}
\caption{\textit{Entanglement phase transition diagrams induced by the non-Hermitian skin effect and its extensions.} Panel (a) shows the entanglement phase diagram of the symplectic Hatano-Nelson model, as studied in Ref.~\cite{nhskin}. Panels (b)–(e) present the corresponding steady-state bipartite entanglement phase diagrams for its variants under disorder~\cite{lzc2}, quasiperiodic potentials~\cite{zhou1}, and the Wannier-Stark ladder~\cite{Li2}, respectively.}
\label{fig:non-Hermitian}
\end{figure}

Non-Hermitian physics exhibits many counterintuitive phenomena that are absent in Hermitian systems, and one of the most intriguing is the non-Hermitian skin effect. It was first introduced in single-particle models to describe an anomalous bulk-boundary correspondence in open-boundary non-Hermitian models with non-reciprocal hopping, commonly known as the Hatano-Nelson model~\cite{hn1,hn2,hn3}. In these models, the majority of eigenstates become unusually localized at one edge. Ref.~\cite{nhskin} first discussed entanglement phase transitions in free-fermion systems induced by the many-body non-Hermitian skin effect. Under open boundary conditions, the nontrivial particle current within the system drives the quench dynamics into a state of non-equilibrium relaxation, ultimately reaching a steady state characterized by an area law, a signature that is directly attributed to the non-Hermitian skin effect, see Fig.~\ref{fig:non-Hermitian} panel (a). It is noteworthy that, even though post-selection removes the randomness introduced by measurement, the non-reciprocity embedded in the effective Hamiltonian still introduces non-unitarity into the dynamics. As a result, a steady-state entanglement phase transition, from a volume-law phase to an area-law phase, emerges as a function of the measurement strength. Furthermore, the findings indicate that the skin effect is a critical factor in triggering the entanglement phase transition. In contrast to measurement-induced entanglement transitions, this criticality is described by an effective central charge that is sensitive to the boundaries, which implies that such behavior may be more widely observable in open quantum systems. The skin effect fundamentally alters the properties of the non-equilibrium steady state by increasing purity and reducing the von Neumann entropy. Subsequent studies have explored entanglement phase transitions induced by the non-Hermitian skin effect in various settings.

In this non-Hermitian skin effect induced entanglement phase transition, a key finding is that under open boundary conditions a stable non-reciprocal particle current exists. This result provides an effective understanding of the non-equilibrium entanglement relaxation from nontrivial initial configurations. In fact, local on-site potentials influence this non-reciprocal particle current. Ref.~\cite{lzc2} numerically studied the entanglement dynamics in the open-boundary disordered Hatano-Nelson free fermion model and found that the introduction of disorder enriches the steady-state entanglement phase diagram of the original Hatano-Nelson model. Specifically, disorder gives rise to a disorder-induced area-law phase, and between this phase and the area-law phase caused by the non-Hermitian skin effect there exists a small critical region characterized by algebraic scaling of entanglement, see Fig.~\ref{fig:non-Hermitian} panel (b). The presence of this critical region suggests that non-reciprocal transport driven by the skin effect competes with Anderson localization induced by disorder, and the associated critical behavior does not display universal invariance. Moreover, because the skin effect is sensitive to boundary conditions, under periodic boundary conditions the skin effect induced area-law phase does not occur. Instead, the entanglement transition exhibits a monotonic localization-induced behavior. This picture is not limited to disordered systems. Ref.~\cite{lsz} also conducted a numerical study of the steady-state entanglement structure under the combined effects of an AAH potential and measurement-induced skin effect. Their findings reveal boundary-dependent area-law phases, an Anderson localization induced area-law phase, and a critical region with algebraic scaling, see Fig.~\ref{fig:non-Hermitian} panel (c). They further proposed that the non-equilibrium steady-state entanglement behavior can be understood within a non-Hermitian single-particle framework. In addition, Ref.~\cite{zhou1} demonstrated that compared to the bidirectional hopping non-reciprocal model discussed in Ref.~\cite{lsz}, a non-Hermitian free fermion system with unidirectional hopping and a quasiperiodic potential exhibits significantly different steady-state entanglement dynamics. In this system, the steady state features an area-law phase induced by Anderson localization and a logarithmic-law phase induced by non-Hermitian effects, with a volume-law scaling region sandwiched between them, see  Fig.~\ref{fig:non-Hermitian} panel (d). This clearly indicates that the steady-state entanglement behavior and critical properties largely depend on the competition between the non-unitary dynamics induced by different types of non-reciprocity in the effective Hamiltonian constructed through measurement operators and the intrinsic unitary dynamics. Similarly, Ref.~\cite{Li2} also investigated the fate of steady-state entanglement in free fermion systems with the non-Hermitian skin effect and Wannier-Stark ladders. Their phase diagrams under open boundary conditions were similar to those reported in Refs.~\cite{lzc2,lsz}, see Fig.~\ref{fig:non-Hermitian} panel (e). Furthermore, finite-size scaling analysis indicates that the entanglement phase transitions in these systems with local potentials do not belong to any known universality class.

\subsection{Entanglement Transitions beyond Non-Hermitian Skin Effects}

The previous framework for steady-state entanglement phase transitions is based on the non-Hermitian skin effect, which originates from the point-gap topology of the system's complex spectrum characterized by a nonzero winding number. This non-Hermitian dynamic property demonstrates that the spectral structure directly determines the distribution of entanglement in quantum states, thereby governing their evolution. In principle, by manipulating measurement operators to engineer different spectral structures, one can achieve entanglement phase transitions, including those associated with the non-Hermitian skin effect, and consequently alter the behavior of steady-state entanglement.

\begin{figure}[tb]
\begin{center}
\includegraphics[width=11.5cm]{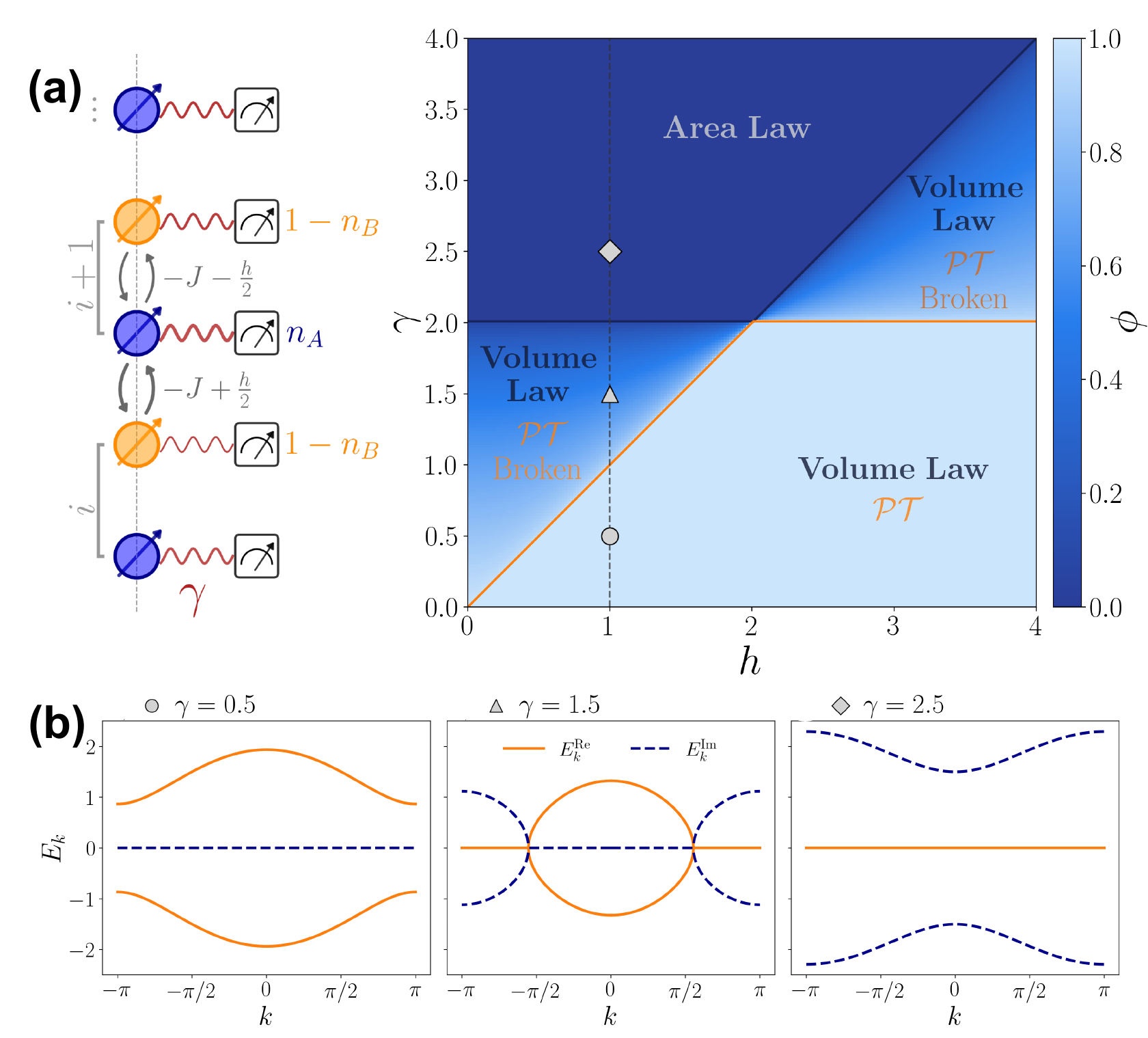}
\end{center}
\caption{\textit{Entanglement phase transition and $\mathcal{PT}$ transition in non-Hermitian SSH model in Ref.~\cite{pointgap}.} Panel (a) displays the phase diagram for the entanglement phase transition and the $\mathcal{PT}$ transition in the non-Hermitian SSH model, highlighting that while the two transitions share similar features, they are not perfectly aligned. Panel (b) depicts the evolution of the energy spectrum as the measurement strength $\gamma$ increases, showing that the system evolves from a $\mathcal{PT}$-symmetric phase to a partially broken phase and finally to a completely broken phase, with significant spectral changes occurring during these transitions.}
\label{fig:spectrum}
\end{figure}

In certain $\mathcal{PT}$-symmetry breaking non-Hermitian models, the reality or complexity of the spectrum is controlled by $\mathcal{PT}$ symmetry. An interesting question arises regarding the relationship between $\mathcal{PT}$ symmetry breaking and entanglement phase transitions in non-Hermitian evolution. Ref.~\cite{pointgap} addressed this issue by studying the non-Hermitian entanglement dynamics of a free fermion chain based on the Su-Schrieffer-Heeger (SSH) model under continuous measurements. Their findings indicate that such systems undergo an entanglement phase transition from volume-law to area-law scaling as the measurement dissipation strength is varied, see Fig.~\ref{fig:spectrum}. By analytically computing the correlation matrix, they characterized both the entanglement dynamics and the steady-state scaling. The transition is fundamentally related to the spectral structure of the system because the quasiparticles acquire a finite lifetime due to measurement backaction, which prevents them from contributing to the volume-law scaling. Fig.~\ref{fig:spectrum}, panel (a), illustrates that the $\mathcal{PT}$-symmetric phase diagram and the entanglement phase diagram share a similar connection in their phase transitions. Panel (b) of Fig.~\ref{fig:spectrum} shows that as the measurement strength $\gamma$ increases, the system evolves through three stages: a $\mathcal{PT}$-symmetric phase, a partially broken phase, and a completely broken phase. These stages are accompanied by significant changes in the energy spectrum. Interestingly, the spectral transition induced by spontaneous $\mathcal{PT}$ symmetry breaking does not exactly coincide with the critical point of the entanglement phase transition. Only when a gap appears in the imaginary part of the spectrum does the steady state truly reach the area-law.

This discussion is further developed in Ref.~\cite{zhou1}, where the authors considered a quasiperiodic free fermion chain with spontaneous $\mathcal{PT}$ symmetry breaking. They discovered that the competition between the coherent hopping strength and the dissipative quasiperiodic potential strength produces a rich entanglement phase diagram. In particular, when the two strengths are comparable, a logarithmic entanglement growth phase emerges. When the dissipative quasiperiodic potential strength exceeds the hopping strength, the combination of the spectral point-gap and bulk localization suppresses entanglement propagation, leading to a steady-state area-law phase. In contrast, when the hopping dominates, a high-entanglement volume-law phase is observed. Moreover, control over the spectrum is not limited to static potential tuning but can also be achieved through periodic Floquet driving, which gives rise to new steady-state entanglement phase transition behaviors. For example, Ref.~\cite{zhou2} constructed a non-Hermitian SSH Hamiltonian with gain-loss Floquet hopping dynamics using measurement operators. Their study revealed that as the non-Hermitian dissipation, or measurement strength, increases, non-monotonic changes in the Floquet spectrum trigger an anomalous reentrant entanglement phase transition. This is manifested by a periodic alternation between volume-law and area-law phases, as the Floquet gap in the imaginary spectrum opens and closes cyclically.

\section{Experiments}\label{sec:5}
Although the MIET was initially proposed and verified mainly in Clifford circuits and generic qubit systems, an increasing number of studies are now focusing on its potential realization in free fermion systems. Owing to the integrability and analytical solvability of free fermion models, along with their well-defined entanglement structure, these systems offer an ideal platform for experimentally observing MIET. Several controlled quantum platforms, including photonic quantum circuits~\cite{photon1,photon2,photon3}, ultracold atomic optical lattices~\cite{rydbergatom}, trapped-ions~\cite{sy1}, and superconducting qubit system~\cite{postselection4}, are attempting to engineer dynamics that approximate free fermion behavior while incorporating tunable measurement mechanisms, thereby driving the experimental realization of MIET.

In photonic systems, researchers leverage linear optical interference and multiphoton entanglement to construct effective simulators for non-interacting fermions~\cite{panjianwei}. Building on this, programmable projection measurement strategies, such as partial path erasure or polarization projection~\cite{projection1,projection2}, have been employed to study how the system's entanglement entropy varies with the measurement rate. Although the scalability of current photonic platforms is limited, these experiments offer important proof-of-concept pathways for realizing MIET in free fermion systems. By tuning the interaction strength between atoms, for example, setting $U=0$ in the Hubbard model, these platforms can precisely simulate free fermion dynamics. Coupled with quantum gas microscopy~\cite{Microscope1,Microscope2}, which enables the readout of individual atomic positions and spin states, the necessary conditions for performing local measurements are met. Some proposals~\cite{learning1,learning2,learning3,leaning4} suggest alternating between Hamiltonian evolution and time-resolved measurements to mimic random measurement circuit dynamics, with the aim of reconstructing the evolution of entanglement entropy or mutual information from experimental data thus indirectly capturing the behavior of MIET. However, these proposals remain in early stages of validation, with one of the central challenges being the post-selection problem~\cite{postselection1,postselection2,postselection3,postselection4}.

Specifically, to observe the entanglement evolution along particular trajectories, experiments often must filter out measurement outcomes that meet certain criteria from a large number of repetitions, a process known as post-selection. This approach faces two major difficulties: first, the probability of obtaining the desired measurement trajectory decreases exponentially with the system size, resulting in very few effective samples; and second, most experimental platforms currently lack non-destructive trajectory readout capabilities, making high-fidelity data acquisition challenging even when the target trajectory is identified. In free fermion systems, where the entanglement structure is highly sensitive to the sequence of measurements, these issues further heighten the dependency on precise post-selection. To address this bottleneck, some studies have introduced quantum non-demolition  measurements~\cite{qnm1,qnm2,qnm3}, active feedback control mechanisms~\cite{wyp,lzc}, and classical computation-assisted trajectory reconstruction algorithms~\cite{algorithms} to enhance post-selection efficiency. Meanwhile, alternative works suggest using surrogate indicators, such as pseudo-entropy~\cite{someworks1,someworks2} or mutual information~\cite{someworks3}, as experimental signals to observe MIET, thus alleviating the reliance on directly measuring the true entanglement entropy.

In a nutshell, although experimental investigations of MIET in free fermion systems are still in their infancy, the theoretical tractability of free fermion models and the gradual maturation of experimental platforms make these systems a critical bridge between the theory of measurement-induced phase transitions and their practical realization. Looking ahead, advances in measurement control techniques and the increased ability to construct medium-scale systems promise to drive significant breakthroughs in the experimental study of MIET based on free fermion systems.

\section{Outlook}\label{sec:6}
In recent years, numerous efforts have significantly deepened our understanding of the entanglement behavior associated with MIET in various free fermion systems. These studies, covering both the entanglement dynamics driven by measurement-induced randomness and those arising from pure non-Hermitian quench dynamics, have uncovered many intriguing mechanisms and clarified the complex dynamical competition between measurement-induced non-unitary evolution and intrinsic unitary evolution.

Nevertheless, despite these substantial advances, many fundamental questions and controversies remain unresolved, offering exciting avenues for future research. One key issue is whether a true BKT transition exists in monitored free-fermion systems. Numerical simulations, limited by finite system sizes and evolution times, can easily misidentify exponential divergences as power laws, while variations in measurement protocols and replica-continuation schemes modify the marginal operators in the effective field theory, leading some analyses to predict a BKT transition and others to rule it out. Decisive verification therefore demands ultra-large-scale simulations coupled with a rigorous finite-size-scaling analysis of multiple observables and a controlled renormalization gruop (RG) treatment within the replica field theory; the principal obstacles are the sheer computational cost and the theoretical ambiguity that arises when replica continuation and RG flow do not commute. Another pressing challenge is the experimental observation of MIET in free-fermion platforms, where overcoming the post-selection problem~\cite{postselection1,postselection2,postselection3,postselection4} is essential for any practical implementation.

In addition, for systems that include quantum jumps, an important question is whether modifying the measurement protocol, such as adopting a frustrated measurement scheme, can alter the local measurement-induced BKT universality class or even generate a new one. Recent work ~\cite{levy} has made progress in this area, but it remains unclear if these findings can be generalized to all frustrated measurement protocols. Furthermore, the relationship between MIET in monitored free fermion systems and localization phenomena is still not fully understood. Some studies suggest that, in a $d$-dimensional system, the physics of a monitored system is closely related to Anderson localization in disordered systems in $d+1$ dimensions~\cite{MIETanderson}. This connection raises further questions regarding the interplay between MIET and different localization regimes, such as disorder-free localization. Another intriguing possibility is whether a rich steady-state topological phase can be achieved by designing an appropriate measurement protocol. Although Ref.~\cite{measurementonly} has discussed this possibility within measurement-only Clifford quantum circuits, similar investigations in simpler monitored free fermion systems remain unexplored. In addition, while quantum entanglement serves as a primary measure of quantum complexity, recent proposals concerning quantum magic~\cite{magic1,magic2,magic3,xtmagic1,xtmagic2,xtmagic3,xtmagic4,xtmagic5}, which reflects quantum complexity on a higher level, indicate that a magic phase transition may also exist. Understanding the relationship between MIET and quantum magic is therefore an important open question.

On the other hand, there are also many unresolved issues in non-Hermitian evolution without measurement randomness. For instance, it is not yet clear what the entanglement dynamics are in a many-body free fermion system that exhibits a single-particle critical non-Hermitian skin effect~\cite{ch7}, nor whether its unique spectral structure can introduce new steady-state entanglement behavior or novel universality classes. Another direction of inquiry concerns the fate of symmetry restoration in monitored-induced non-Hermitian free fermion systems, including whether phenomena such as the quantum Mpemba effect~\cite{qmp,mpbls,ls_post2,mpbxhek,xtmpb,EAmpb} persist. Recent discussions on the entanglement dynamics and emergent dynamical phase transitions in quantum feedback-induced skin effects~\cite{wyp,Feng_2023,fx,lzc,huang} have suggested that quantum feedback may bridge quantum jump dynamics and non-Hermitian dynamics. However, it remains uncertain whether the skin effect is a necessary condition for these dynamical phase transitions or what mechanisms underlie them.

In summary, although significant progress has been made in understanding MIET in various monitored free fermion systems, many fundamental questions remain open. Continued research, ranging from theoretical investigations to potential technological applications, promises to yield exciting discoveries in the years ahead.

\section*{Acknowledgments}
We would like to thank Shan-Zhong Li, Zhi Li, Ching Hua Lee, Shuo Liu, Ze-Chuan Liu, Xhek Turkeshi, Xuyang Huang and Yu-Jun Zhao for helpful discussions. X.-J. Yu was supported by the National Natural Science Foundation of China (Grant No.12405034) and a start-up grant from Fuzhou University.  J.-X. Zhong was supported the National Natural Science Foundation of China (Grant No.12374046 and No.11874316), the Shanghai Science and Technology Innovation Action Plan (Grant No.24LZ1400800), the National Basic Research Program of China (Grant No.2015CB921103), and the Program for Changjiang Scholars and Innovative Research Teams in Universities (Grant No.IRT13093).

\section*{References}
\bibliographystyle{iopart-num}

\providecommand{\newblock}{}
\bibliography{refs}






\end{document}